\documentclass[conference,a4paper]{IEEEtran}
\IEEEoverridecommandlockouts
\usepackage{cite}
\usepackage[ruled,vlined,linesnumbered]{algorithm2e}
\usepackage{amsmath,amssymb,amsfonts}
\usepackage{algorithmic}
\SetAlCapNameFnt{\footnotesize}
\SetAlCapFnt{\footnotesize}
\SetAlgorithmName{Algorithm}{Algorithm}{List of Algorithms}
\usepackage{graphicx}
\usepackage{stmaryrd}
\usepackage{textcomp}
\usepackage{array}
\usepackage{commath}
\usepackage{sidecap}
\usepackage{stfloats}
\usepackage{tabularx, boldline}
\usepackage{rotating,booktabs,multirow}
\usepackage{mathtools}
\usepackage{flexisym}
\usepackage{breqn}
\usepackage{url}
\usepackage{adjustbox}
\usepackage{makecell}
\usepackage{textcomp}
\usepackage{amsmath}
\usepackage{txfonts}
\usepackage{acronym}
\usepackage{tabularx, boldline}
\usepackage{rotating,booktabs,multirow}
\usepackage{enumerate}
\usepackage{xcolor}
\def\BibTeX{{\rm B\kern-.05em{\sc i\kern-.025em b}\kern-.08em
    T\kern-.1667em\lower.7ex\hbox{E}\kern-.125emX}}

\def\R{\mathbb{R}} 
\def\E{\mathbb{E}} 

\def\LOC{\text{LOC}}
\def\RB{\text{RB}}

\begin{document}

\title{Privacy-Cost Management in Smart Meters Using Deep Reinforcement Learning
}
\author{\IEEEauthorblockN{Mohammadhadi Shateri}
\IEEEauthorblockA{\textit{McGill University} \\  Montreal, Canada}
\and
\IEEEauthorblockN{Francisco Messina}
\IEEEauthorblockA{\textit{McGill University} \\  Montreal, Canada}
\and
\IEEEauthorblockN{Pablo Piantanida}
\IEEEauthorblockA{CentraleSup\'elec-CNRS-Universit\'e Paris Sud \\
 Gif-sur-Yvette, France}
\and
\IEEEauthorblockN{Fabrice Labeau}
\IEEEauthorblockA{\textit{McGill University} \\  Montreal, Canada}}

\maketitle

\begin{abstract}
Smart meters (SMs) play a pivotal rule in the smart grid by being able to report the electricity usage of consumers to the utility provider (UP) almost in real-time. However, this could leak sensitive information about the consumers to the UP or a third-party. Recent works have leveraged the availability of energy storage devices, e.g., a rechargeable battery (RB), in order to provide privacy to the consumers with minimal additional energy cost. In this paper, a privacy-cost management unit (PCMU) is proposed based on a model-free deep reinforcement learning algorithm, called deep double $\mathrm{Q}$-learning (DD$\mathrm{Q}$L). Empirical results evaluated on actual SMs data are presented to compare DD$\mathrm{Q}$L with the state-of-the-art, i.e., classical $\mathrm{Q}$-learning (C$\mathrm{Q}$L). Additionally, the performance of the method is investigated for two concrete cases where attackers aim to infer the actual demand load and the occupancy status of dwellings. Finally, an abstract information-theoretic characterization is provided. 
\end{abstract}


\section{Introduction}
\IEEEPARstart{S}{mart} meters (SMs) play an important rule in the modern electricity network, known as the smart grid (SG), by providing fine-grained power consumption measurements of households to the utility provider (UP) almost in a real-time basis. Although these high-resolution SMs measurements can improve the efficiency, reliability, and flexibility of power infrastructures, they might be used by adversaries or malicious third-parties to violate the privacy of households \cite{giaconi2018privacy}. For instance, by using non-intrusive load monitoring (NILM) approaches, an adversary with access to the power consumption load profiles can readily infer sensitive information related to consumers' daily habits and types of appliances owned. 

To protect consumers' sensitive information, two main families of SM data privacy enabling techniques have been proposed in the literature: i) SM data manipulation methods \cite{sankar2013smart, yang2016evaluation,shateri2019deep}; and ii) demand shaping methods \cite{kalogridis2010privacy,gomez2014smart,giaconi2017smart,sun2017smart,li2018information,erdemir2019privacy}. In the first family, the SM data are processed and manipulated, e.g., by adding random noise, before sharing that with the UP. However, this might cause a significant loss in terms of the utility of the SM data \cite{giaconi2018privacy}. On the other hand, approaches of the second family rely on using physical resources available at the dwelling such as a rechargeable battery (RB), electric vehicles (EVs), heating, ventilation, and air conditioning (HVAC) units, renewable energy source (RES), or a combination of them, to shape the way that energy is consumed by the user from the grid. In this case, the goal is to find an optimum energy management strategy, under the physical limitations of the resources, which provides maximum privacy with minimal household energy expenses.

In \cite{li2018information}, the problem of finding an optimum strategy was formulated as a Markov decision process (MDP) and solved using dynamic programming. Concretely, a single-letter expression of the minimum information leakage was provided for the case that demand load is independent and identically distributed (i.i.d.). However, this result is mostly of theoretical value since the demand load is not i.i.d. in practice. In addition, the energy cost was not included in this problem formulation. In another study, the joint optimization of privacy, measured by fluctuations of the grid load around a constant load, and energy cost was considered and modeled as an MDP \cite{sun2017smart}. To solve this MDP, a classical model-free algorithm from reinforcement learning called $\mathrm{Q}$-learning \cite{sutton2018reinforcement} was used. This tabular algorithm iteratively learns the optimal state-action value function $\mathrm{Q}^*$ from which an optimal policy for using the physical resources is readily obtained. Nevertheless, the classical $\mathrm{Q}$-learning (C$\mathrm{Q}$L) convergence properties can be poor when the state and/or action spaces are large or infinite. 

In this paper, we adopt the MDP formulation introduced in \cite{sun2017smart} and use a deep learning approximation of the C$\mathrm{Q}$L algorithm called deep double $\mathrm{Q}$-learning (DD$\mathrm{Q}$L) to overcome the previously discussed limitation of the classical algorithm. To the best of our knowledge, this is the first work that uses deep reinforcement learning to design a privacy-preserving mechanism for smart meters. For the sake of simplicity, we only consider an RB and assume that no energy can be sold to the grid (however, the problem formulation can be easily extended to different scenarios). In addition, the formulation of the cost is revised to take into account not only the total electricity cost but also the cost associated with the wear and tear of the battery. The performance of the DDQL algorithm is assessed using actual SMs data both from practical and theoretical perspectives. For the former evaluation, we consider two scenarios modeling different privacy threats. In the first scenario, an attacker observes the grid load and attempts to infer the consumer's demand load; while in the second scenario an attacker is trained to infer the occupancy status of the household. The latter evaluation is done by studying the trade-off between the cost and the mutual information between the demand and grid loads.

\section{Problem Formulation}

\subsection{Demand Shaping Using Rechargeable Battery}

\begin{figure}[htbp]
	\centering
	\includegraphics[width=0.7\linewidth]{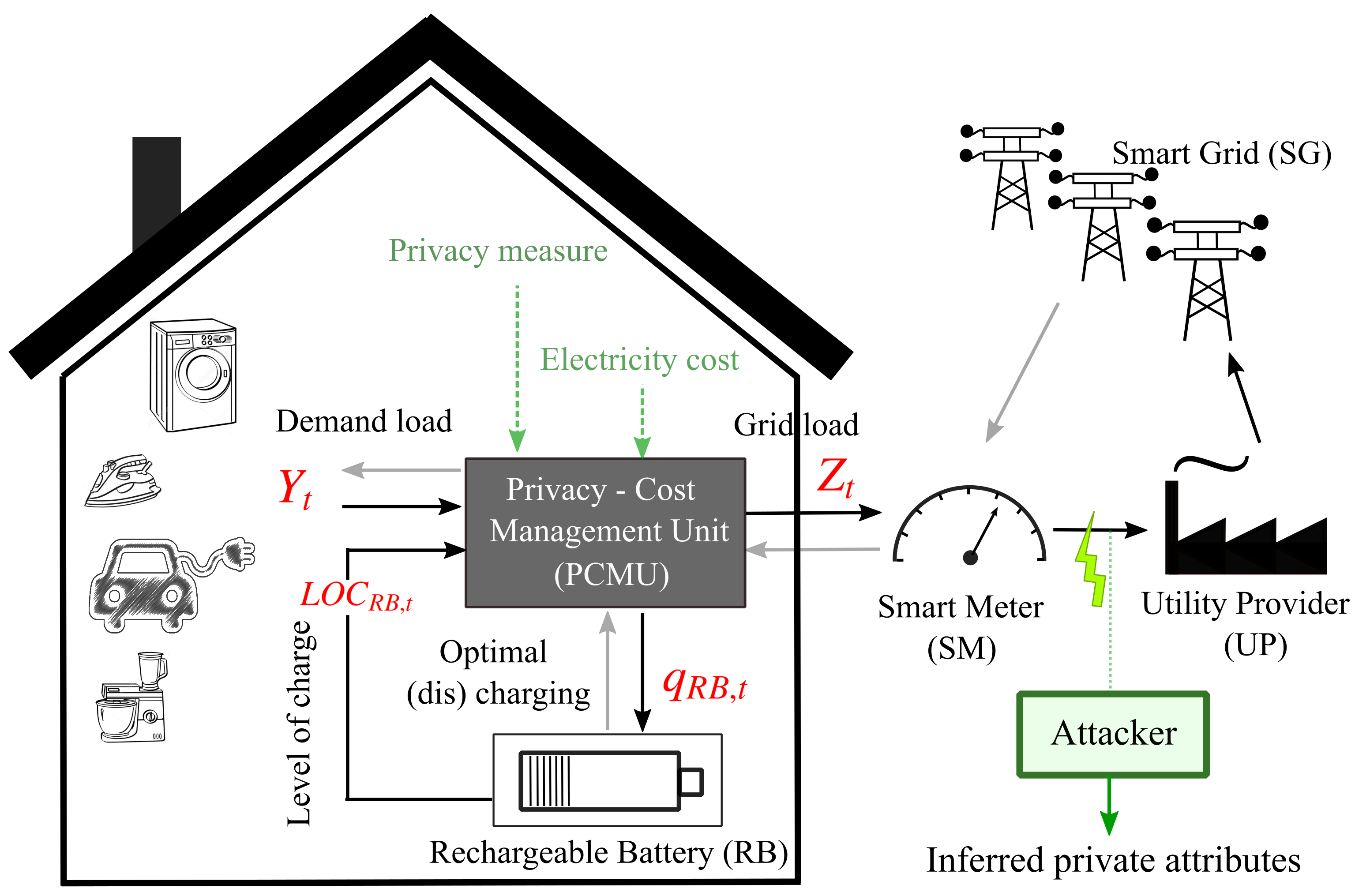}
	\caption{Privacy-preserving model framework for smart meters based on a rechargeable battery.}
	\label{RB-Based Model}
\end{figure}

Fig. \ref{RB-Based Model} shows the privacy-preserving framework for a household based on a single rechargeable battery (RB). Here, the attacker can be the UP itself, an eavesdropper or a third-party. In this framework, a privacy-cost management unit (PCMU) is designed to determine the optimal charging/discharging rate of the battery with the purpose of masking the consumer's demand load while minimizing the total electricity cost.

Let $y_t$ be the consumer's demand load/power at time $t$, where $t \in \mathcal{T} =\{ 1, \ldots, T \}$. Given the demand load $y_t$ and the level of charge of battery $\LOC_{RB,t} \in [0,1]$ at time $t$, the PCMU should determine the optimum charging/discharging rate $q_{RB,t}$ to hide the consumer's demand load. Therefore, the SM  measures and reports the masked demand load (i.e., the grid load) given by
\begin{equation} \label{eq:Releaseload}  z_t = y_t + q_{RB,t}. \end{equation}
The PCMU should be able to limit the performance of a potential attacker trying to infer sensitive information (which could be either $y_t$ or a correlated variable) from $z_t$. In this framework, there is a trade-off between privacy and cost associated to energy and wear and tear of the battery.

\subsection{Markov Decision Process (MDP) Model}

The problem of finding an optimal policy for the PCMU (i.e., the reinforcement learning agent in our problem) can be formalized using the framework of Markov decision processes (MDPs) \cite{sutton2018reinforcement}, as in \cite{sun2017smart,li2018information,erdemir2019privacy}. An MDP is determined by a state space $\mathcal{S}$, which defines the possible states; an action space $\mathcal{A}(s)$ for each $s \in \mathcal{S}$, which describes the feasible actions at state $s$ (we use $\mathcal{A}$ to denote $\cup_{s\in\mathcal{S}} \; \mathcal{A}(s)$); the environment dynamics $p(s_{t+1}|s_t,a_t)$, which dictates how the current state $s_t$ changes when action $a_t$ is taken; and the reward function $r(s_t,a_t)$ which gives the immediate reward received at state $s_t$ by taking action $a_t$; and the discount factor $\gamma \in [0,1]$ which determines the decay rate for weighting future rewards. In this study, we consider a finite horizon time model, i.e. we assume $T<\infty$. Starting from an initial state $s_1$, and by following a policy $\pi(a|s) = p(a|s)$, the PCMU takes an action $a_1$, receives a reward $r_1$, and transitions to state $s_2$. This leads to generating a sequence of states, actions and rewards: $\left[s_1,a_1,r_1,\ldots,s_T,a_T,r_T\right]$, which is referred to as an episode. The action-value function $\mathrm{Q} : \mathcal{S} \times \mathcal{A} \to \R$ is defined as $\mathrm{Q}(s,a) = \E_{\pi}\left[\sum_{t=1}^{T} \gamma^{t-1} r\left(s_t,a_t\right) \bigg |s_1=s,a_1=a\right]$ and determines how good is to take action $a \in \mathcal{A}(s)$ in state $s \in \mathcal{S}$ and then following the policy $\pi$. An optimal policy $\pi^*$ is one that maximizes $\mathrm{Q}(s,a)$ for all possible $s$ and $a$. In our problem, the state is defined as $s_t = [\LOC_{RB,t}, y_t]^T$ while the action is defined as the RB charging/discharging rate, i.e. $a_t = q_{RB,t}$, where a positive $q_{RB,t}$ indicates the RB is charging and a negative $q_{RB,t}$ means discharging of RB for supplying the consumer's demand. In addition, the environment transition probability $p(s_{t+1}|s_t, a_t)$ can be written as follows:
\begin{equation} \label{eq:MDP_Transition} 
p\left(s_{t+1}|s_t, a_t\right) \overset{\text{(i)}}{=} p\left(\LOC_{RB,t+1}|\LOC_{RB,t},q_{RB,t}\right)\times p\left(y_{t+1}|y_t\right)
\end{equation}
where (i) is due to the assumption that the consumer's demand load is independent of the action and level of charge of battery. The factor $\small{p\left(\LOC_{RB,t+1}|\LOC_{RB,t},q_{RB,t}\right)}$ is determined by the dynamics and physical constraints of the battery, which can be summarized as follows \cite{sun2017smart}: 
\begin{equation} \label{eq:ChargingConstraint} 
q_{\RB}^{\min}\leq q_{\RB,t}\leq q_{\RB}^{\max},
\end{equation}
\begin{equation} \label{eq:LOC-Dynamics} 
\LOC_{\RB,t+1} = \LOC_{\RB,t} + \frac{q_{\RB,t}\times \Delta t\times e_{\RB}}{C_{\RB}},
\end{equation}
\begin{equation} \label{eq:LOCConstraint} 
\LOC_{\RB}^{\min}\leq \LOC_{\RB,t}\leq \LOC_{\RB}^{\max},
\end{equation}
where $q_{\RB}^{\min}$ and $q_{\RB}^{\max}$ are the minimum and maximum charging rate of the RB, $\LOC_{\RB}^{\min}$ and $\LOC_{\RB}^{\max}$ are the minimum and maximum level of charge of the RB, $\Delta t$ is the load sampling rate, $e_{\RB}$ is the charging efficiency factor of the RB, and $C_{\RB}$ is the capacity of the RB. However, note that $p\left(y_{t+1}|y_t\right)$ is unknown and in general difficult to approximate \cite{sun2017smart}. Thus, we do not assume full knowledge of the environment dynamics.

It remains to determine the reward function, which can inversely be interpreted as a loss function: $\ell\left(s_t,a_t\right) = -r\left(s_t,a_t\right)$. Following \cite{sun2017smart}, the loss function will be defined as a convex combination of a privacy measure term and an electricity cost term. In particular, privacy is defined as the distance between the reported load by the SM and a (predetermined) flat load. Intuitively, when (regardless of the consumer's demand) a constant load is reported by the SM, we expect that no private information about the consumer is leaked. Concretely, considering $l_c$ as the desired constant load, and following policy $\pi$, privacy is defined as the average relative distance of the grid load from the target load as follows:
\begin{equation} \label{eq:PrivacyMeasure} 
F_{\pi} := \frac{1}{T}\sum_{t=1}^{T} f\left(s_t,a_t\right)=\frac{1}{T}\sum_{t=1}^{T} \left |\frac{z_t - l_c}{l_c} \right |.
\end{equation}
It should be noted that other notions of privacy have also been considered for this problem \cite{giaconi2018privacy}. On the other hand, the cost that is incurred by the user can be related to the electricity cost and also to the cost associated to the battery wear and tear. In this study, we assume that no energy can be sold to the grid. Therefore, the average electricity cost over time horizon $T$ is computed as follows:
\begin{equation} \label{eq:Costaverage} 
C_{\pi} = \frac{1}{T}\sum_{t=1}^{T} \Delta t\times h_t \times \left [ z_{t} \right ]^{+},
\end{equation}
where $h_t$ is the price of $1$ kWh of energy purchased from grid at time $t$ and $[x]^+ = \max(x,0)$. Since $\left [ z_{t} \right ]^{+} = \left [ y_{t} +q_{\RB,t} \right ]^{+}\leq  y_{t} +\left |q_{\RB,t} \right |$, and the PCMU has no control on $y_{t}$, we conclude that minimizing the average additional cost
\begin{equation} \label{eq:CostMeasure} 
G_{\pi} := \frac{1}{T}\sum_{t=1}^{T} g(s_t,a_t) = \frac{1}{T}\sum_{t=1}^{T} \Delta t\times h_t \times \left | q_{\RB,t} \right |, \end{equation}
minimizes an upper bound of $C_{\pi}$. In addition, notice that $G_{\pi}$ takes into account the battery wear and tear cost since it grows when the battery use increases. Finally, considering equations \eqref{eq:PrivacyMeasure} and \eqref{eq:CostMeasure}, the one-step loss function is defined as follows:
\begin{equation} \label{eq:CostFunction} 
\ell\left(s_t,a_t\right) = -r\left(s_t,a_t\right) = \lambda g(s_t,a_t) + (1-\lambda) f(s_t,a_t),
\end{equation}
where $\lambda \in [0,1]$ controls the privacy-cost trade-off. 




\section{Methodology}

\subsection{Classical $\mathrm{Q}$-Learning (C$\mathrm{Q}$L) Algorithm}


The $\mathrm{Q}$-Learning ($\mathrm{Q}$L) algorithm \cite{sutton2018reinforcement} is a simple method to learn the optimal state-action value function $\mathrm{Q}^*$. Then, the best action at state $s$ is readily obtained by maximizing $\mathrm{Q}^*(s,\cdot)$. In the process of learning, the Q-Learning algorithm takes an action $a_t$ at state $s_t$ using some policy, transitions to state $s_{t+1}$ and updates the $\mathrm{Q}$ function as follows\cite{sutton2018reinforcement}:
\begin{align} \label{eq:QLearning}
\Delta Q\left(s_t,a_t\right) = \alpha \; \bigg( r\left(s_t,a_t\right) + \gamma \underset{a\in \mathcal{A}\left(s_{t+1}\right)}{\text{max}} Q\left(s_{t+1},a\right) - Q\left(s_t,a_t\right)\bigg), \end{align}
where $\alpha \in [0,1]$ is the learning rate, which should be selected wisely for the sake of convergence \cite{szepesvari1998asymptotic}. In general, the $\mathrm{Q}$-Learning algorithm uses the $\epsilon-$greedy policy, in which a random action is taken with probability $\epsilon$, and the action which maximizes the $\mathrm{Q}$-value is taken with probability $1 - \epsilon$. In this way, while the algorithm exploits the learned $\mathrm{Q}$ function, it also explores the action space to discover the benefits of taking other actions (this is known as the exploration-exploration dilemma). Therefore, starting from an initial state and initializing the $\mathrm{Q}$ function/table to some arbitrary values, we take an action using the $\epsilon-$greedy policy to move to the next state and update the $\mathrm{Q}$ function based on \eqref{eq:QLearning}. This is continued for all the next states until the end of episode. The process is then repeated for several episodes to ensure all the state-actions are visited enough times. 

\subsection{Deep Double $\mathrm{Q}$-Learning (DD$\mathrm{Q}$L) Algorithm}

One main issue regarding the classical $\mathrm{Q}$-Learning (C$\mathrm{Q}$L) algorithm is that it needs to visit all the state-action pairs several times to provide a good approximation of $\mathrm{Q}^*$. Therefore, for large MDPs with many states and actions, convergence is very slow. To resolve this problem, the Q-function can be approximated by using a deep neural network (DNN). These new methods, where deep learning is used for approximating the Q-function, are called deep $\mathrm{Q}$-Learning (D$\mathrm{Q}$L) methods \cite{volodymyr2015human,franccois2018introduction}. Recently, the D$\mathrm{Q}$L method was used for household energy management in \cite{mocanu2018line}. The main idea of the D$\mathrm{Q}$L method studied here is to approximate $\mathrm{Q}^{*}(s,a)\approx \mathrm{Q}(s,a;\theta)$ where $\theta$ are the parameters of a DNN called the $\mathrm{Q}$-Network. The Q-Network takes the state $s\in\mathcal{S}$ at the input and generates $\mathrm{Q}(s,a)$ at the output for different actions $a\in\mathcal{A}$. To define the objective function for this $\mathrm{Q}$-network, we observe from \eqref{eq:QLearning} that convergence is obtained when the term in parenthesis is equal to zero. The term $r\left(s_t,a_t\right) + \gamma \; \text{max}_{a\in \mathcal{A}(s_{t+1})} \mathrm{Q}\left(s_{t+1},a;\theta\right)$ can be interpreted as the target, while $\mathrm{Q}\left(s_t,a_t;\theta\right)$ is the output of the Q-Network. Thus, the mean squared error loss between target and output can be used as the loss function for training the $\mathrm{Q}$-Network. However, using the same network to compute the target and output often leads to instability \cite{mnih2015human}. To address this issue, the so-called double Q-learning algorithm was proposed in \cite{hasselt2010double} and extended to the deep learning setting in \cite{van2016deep}. In the DDQL algorithm, a second network called the Target Network (with parameters $\theta^{\prime}$) is used to calculate the target term. The Target Network parameters $\theta^{\prime}$ are periodically updated by simply copying the parameters from the Q-Network. Thus, using the Target Network, the objective function of the $\mathrm{Q}$-Network can be written as follows:
\begin{align} \label{eq:DDQL_loss}
\mathcal{L}_{QN}(\theta) =& \E\bigg[ \bigg(r(s_t,a_t) +
\gamma \underset{a\in \mathcal{A}(s_{t+1})}{\text{max}} Q(s_{t+1},a;\theta^{\prime})-Q(s_t,a_t;\theta) \bigg)^2\bigg].
\end{align}
The training of the DD$\mathrm{Q}$L is presented in Algorithm \ref{AlDDQL}.




\begin{algorithm}
    \footnotesize
    \algsetup{linenosize=\tiny}
	\caption{Training of the deep double $\mathrm{Q}$-learning algorithm. Copy step $k$ and training step $k^{\prime}$ are hyperparameters.}
	\label{AlDDQL}
	\begin{algorithmic}[1]
	    \STATE Initialize $\mathrm{Q}$-network and target network.
		\FOR {number of training episodes}
		\STATE Set the initial state $s_1=[\LOC_{\RB,1},y_1]$.
		\FOR {$t=1,\ldots,T$}
		\STATE Observe the state $s_t=[\LOC_{\RB,t},y_t]$
		\STATE Select feasible action $a_t$ using $\epsilon-$greedy algorithm.
		\STATE Calculate reward $r(s_t,a_t)$ from equation \eqref{eq:CostFunction}. 
		\STATE Update the next state $s_{t+1}$ based on \eqref{eq:LOC-Dynamics} and \eqref{eq:LOCConstraint}.
		\STATE Import $(s_t,a_t,r(s_t,a_t),s_{t+1})$ into the replay buffer.
		\STATE Every $k^{\prime}$ step, update the $\mathrm{Q}$-network by minimizing \eqref{eq:DDQL_loss} using samples from the replay buffer.
		\STATE Every $k$ step, update the target network by copying the\\ $\mathrm{Q}$-network parameters.
		\ENDFOR
		\ENDFOR
	\end{algorithmic}
\end{algorithm}

\section{Results and Discussion} \label{sec:results}

\subsection{Description of data set and parameters} \label{sec:dataset}
In this study the electricity consumption and occupancy (ECO) data set published by \cite{beckel2014eco} is used. This data set includes 1 Hz electricity usage measured by smart meters and occupancy labels gathered through a tablet computer and a passive infrared sensor. In this study, the data set is sampled with a sampling rate $\Delta t=15$ min and an episode is defined over a day. Totally $2700$ samples (each a vector with length of 96) are used and split into training, validation, and test sets with ratio \mbox{70:10:20}, respectively. The validation dataset is used to set the values of the hyperparameters for each algorithm, which are discussed in the next section. The desired constant load is $l_c = 0.7$ kW and the parameters of the battery are as follows: $C_{\RB} = 10$ kWh, $e_{\RB}=1$, $q_{\RB}^{\max}=-q_{\RB}^{\min}=4$ kW, $\LOC_{RB}^{\max}=1$ and $\LOC_{\RB}^{\min}=0$. We consider the winter Time-of-Use tariff offered by Ontario/Canada, in which the off-peak price is $\$0.101$ kWh during 19:00 to 7:00, the mid-peak price is $\$ 0.144$ kWh during 11:00 to 17:00, and the on-peak price is $\$0.208$ kWh during 7:00 to 11:00 and during 17:00 to 19:00.

\subsection{Deep double $\mathrm{Q}$-learning  versus Classical $\mathrm{Q}$-learning } \label{sec:DQLCQL}

In this section, the performance of the DD$\mathrm{Q}$L method is empirically compared with C$\mathrm{Q}$L. For both methods, the network configuration and the value of the hyperparameters are determined empirically by the best privacy-cost trade-off over the validation dataset. 

Considering the tabular setting of the C$\mathrm{Q}$L algorithm, the state and action states need to be quantized. To this end, a total number of 160 and 100 quantization levels were used for action and demand load, respectively. Therefore, according to the equation \eqref{eq:LOC-Dynamics} and our definition of state $s_t$ the $\mathrm{Q}$ function would be a table with size $800\times100\times160$. A total number of 25K episodes ($\approx$ 2.5M steps) are used with a discount factor $\gamma = 0.8$ and an adaptive learning rate $\alpha$ decreasing linearly from 0.5 to 0.05 over 1M steps.

On the other hand, in the DD$\mathrm{Q}$L method, a multilayer perceptron (MLP) with two hidden layers, each including 64 neurons and rectified linear unit (ReLU) as activation function, is used for both the Q and Target Networks. A total number of 800 episodes are used and the discount factor $\gamma$ is set as 0.99. The size of the experience replay memory is 10K tuples. The memory gets sampled to update the Q-network every 8 steps ($k^{\prime}$=8), with minibatches of size 128, and a Target Network copy step $k$ of 500 steps. The RMSProp optimizer with learning rate 0.00025 is selected to train the network.

\begin{figure}[htbp]
	\centering
	\includegraphics[width=0.8\linewidth]{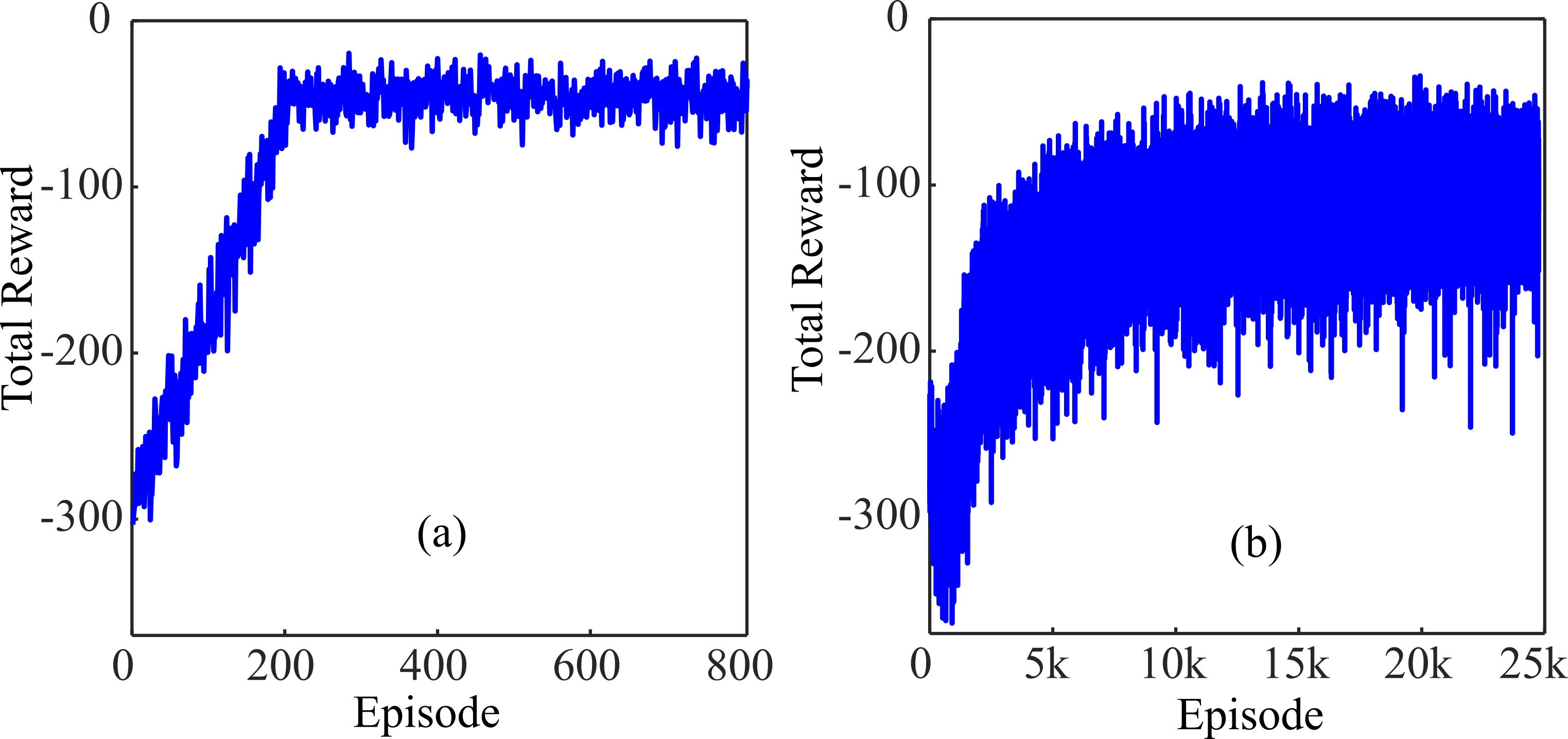}
	\caption{Total episodic reward for (a) DD$\mathrm{Q}$L versus (b) C$\mathrm{Q}$L during training.}
	\label{Reward}
\end{figure}

In the following, the performances of the C$\mathrm{Q}$L and DD$\mathrm{Q}$L algorithms used to train the PCMU unit are evaluated. Fig. \ref{Reward} shows how the total episodic reward evolves during the training of each algorithm (for $\lambda =0$). From this figure, it is clear that, compared with the C$\mathrm{Q}$L, the DD$\mathrm{Q}$L algorithm leads to higher rewards in a much smaller number of episodes.

\begin{figure}[htbp]
	\centering
	\includegraphics[width=0.52\linewidth]{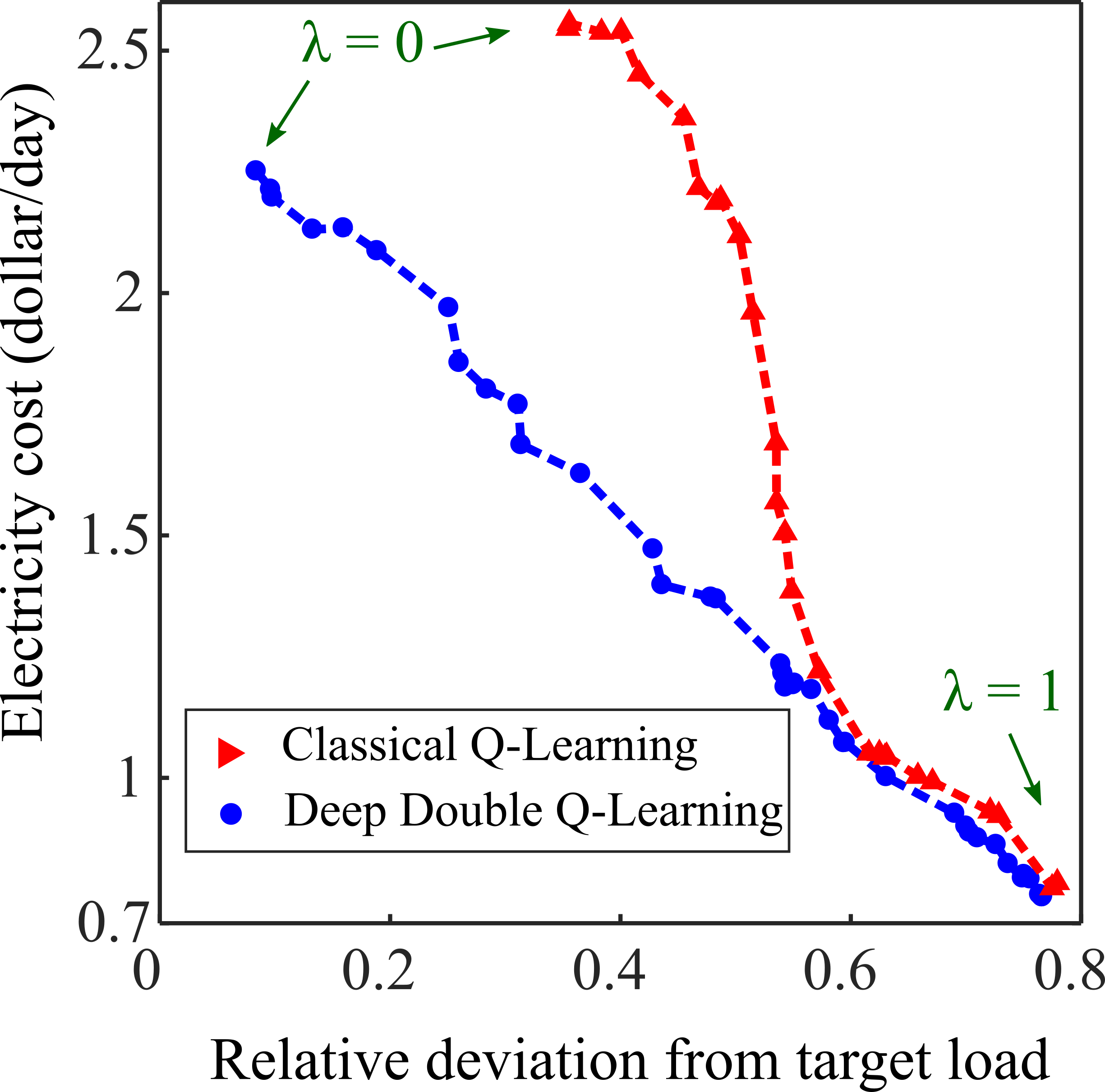}
	\caption{Average daily electricity cost versus average absolute relative deviation from target load (constant load) on test dataset for $\lambda \in [0,1]$.}
	\label{tradeoff}
\end{figure}

To compare the performance of these two algorithms on the test dataset, the trade-off between electricity cost in equation \eqref{eq:Costaverage} and privacy measure in equation \eqref{eq:PrivacyMeasure} are examined for both algorithms in Fig. \ref{tradeoff}. It can be seen that, although for values of $\lambda$ close to 1 both C$\mathrm{Q}$L and DD$\mathrm{Q}$L produce very similar results, for higher privacy levels DD$\mathrm{Q}$L significantly outperforms C$\mathrm{Q}$L. The reason can be understood by looking at Fig. \ref{Reward}, where the DD$\mathrm{Q}$L shows a much better convergence compared with the C$\mathrm{Q}$L. In other words, due to the big state and action spaces, the C$\mathrm{Q}$L algorithm is not able to converge to the same level (even with a much larger number of episodes) and so provides degraded results compared with DD$\mathrm{Q}$L. These results confirm that, in our problem, the DD$\mathrm{Q}$L algorithm can provide better results than the classical C$\mathrm{Q}$L method.

\subsection{Deep double $\mathrm{Q}$-learning versus Attacker } \label{sec:DQLAttacker}

\begin{figure}[htbp]
	\centering
	\includegraphics[width=0.81\linewidth]{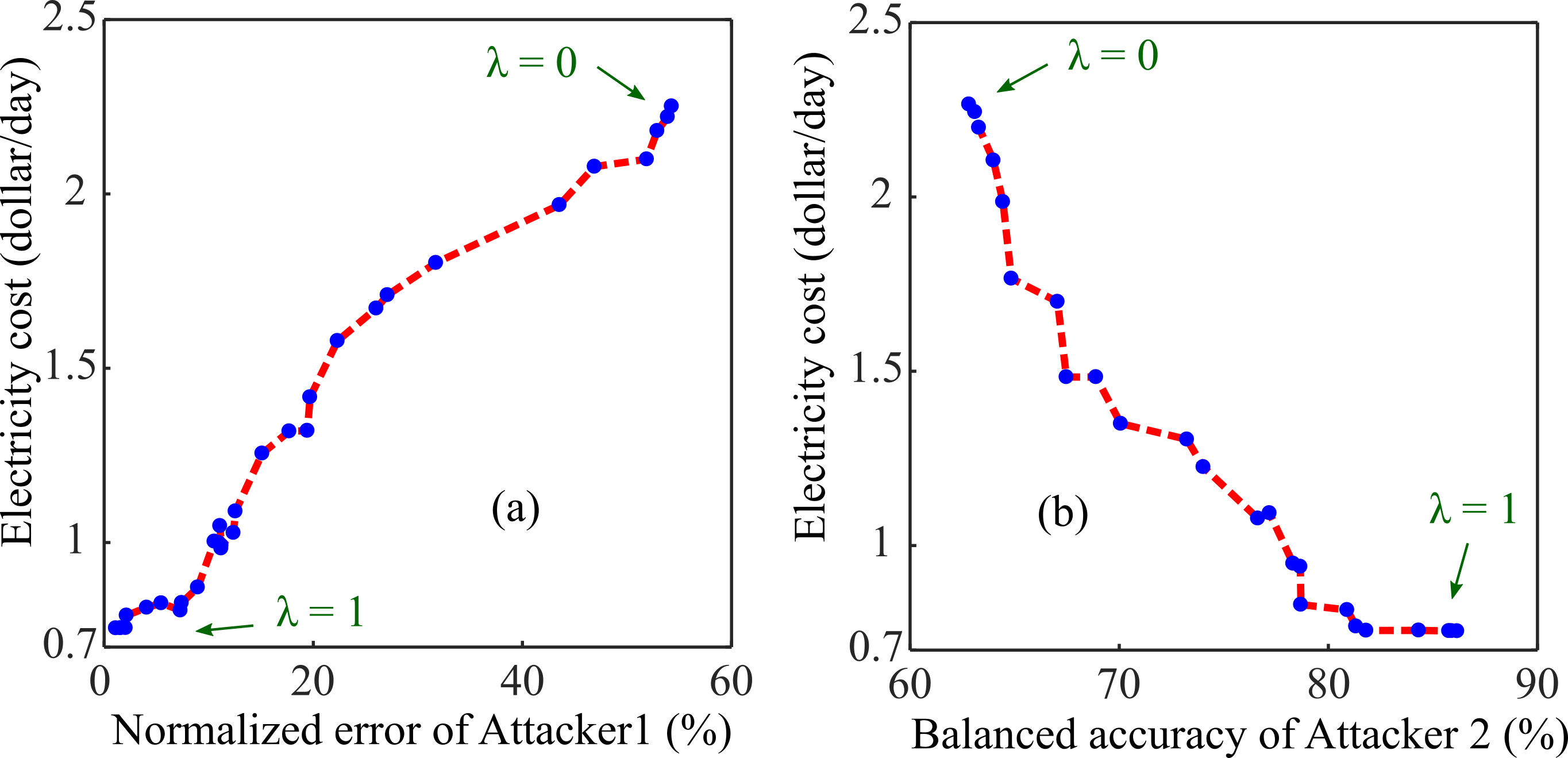}
	\caption{Attackers performances: (a) inferring actual demand load, (b) inferring occupancy status of household.}
	\label{attackers}
\end{figure}

In this section, we evaluate how DD$\mathrm{Q}$L results used in the smart meter privacy-preserving model can reduce attacker inference about user's private attributes. To this end, two scenarios are studied. In the first scenario, an attacker modeled as a neural network with three hidden layers (each with 32 neurons and ReLU activation functions) uses the grid load sequence $z^T$ (with length $T=96$) to infer the user's demand load $y^T$. In the second scenario, an attacker modeled as a neural network with two hidden layers (each with 44 neurons and ReLU activation functions) uses the sequences of grid load $z^T$ to infer the occupancy status of households. Both attackers are given a labeled dataset for training (which is a worst-case assumption), and for both of them the RMSProp optimizer with learning rate $0.001$ is used. The performances of both attackers are presented in Fig. \ref{attackers}. Notice from the results that the PCMU, using the DD$\mathrm{Q}$L algorithm with a RB, can effectively reduce the inference of private information by an attacker by controlling $\lambda$. For example, in the second scenario in which privacy does not matter ($\lambda =1$), the attacker can infer occupancy labels with a balanced accuracy of more than $85\%$, while for $\lambda =0$ it is reduced to less than $65\%$. It should be noted that, by either using a larger battery or other resources, one could in principle approach full privacy (e.g., balanced accuracy of $50 \%$). 

\begin{figure}[htbp]
	\centering
	\includegraphics[width=0.5\linewidth]{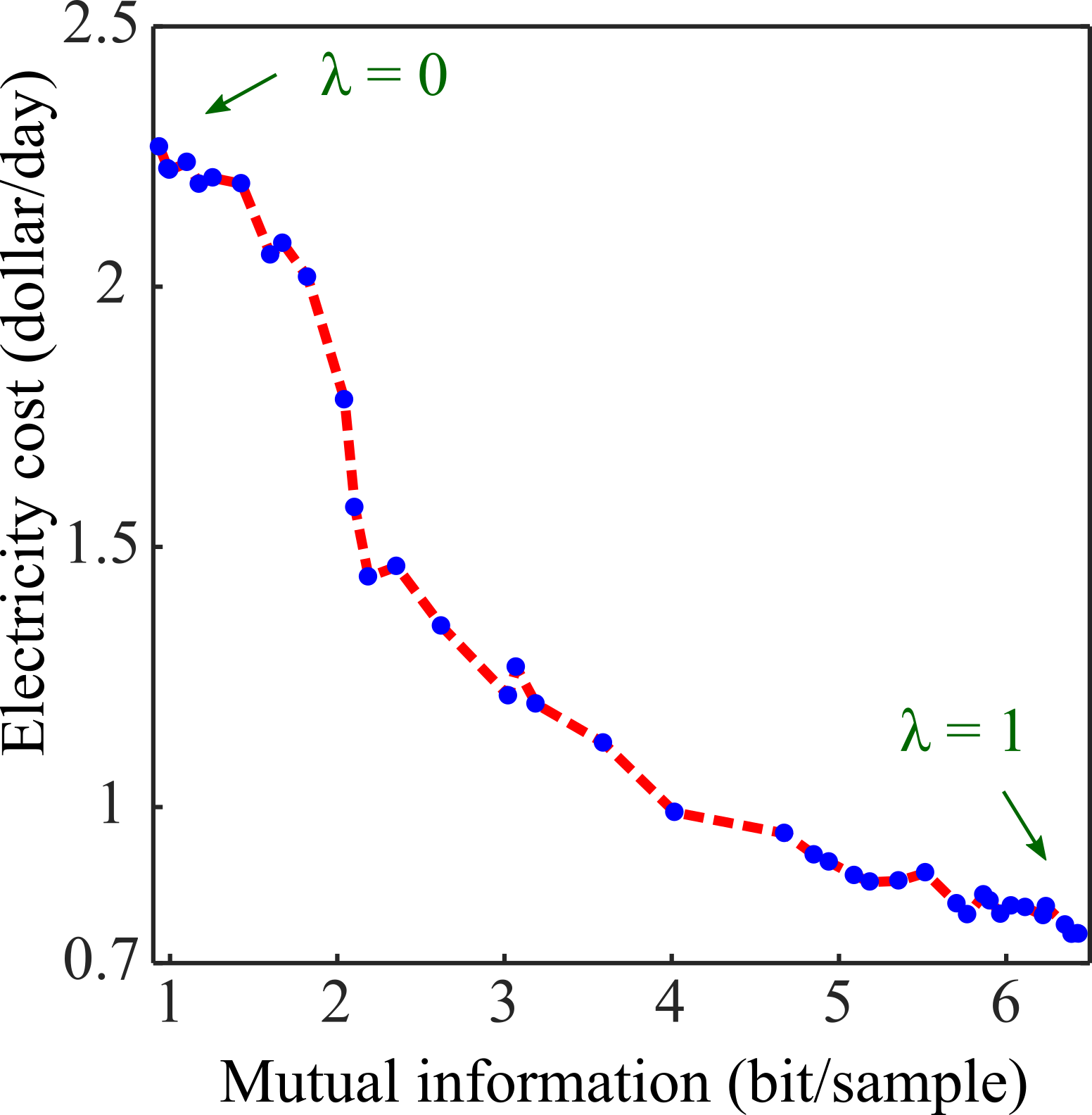}
	\caption{Electricity cost versus MI between demand load and grid load.}
	\label{MI}
\end{figure}

Finally, to investigate the overall leakage of information about the consumers' demand load from the grid load, the mutual information (MI) between demand load and grid load is calculated based on the Kraskov–Stögbauer–Grassberger (KSG) estimation method (with 4 neighbors) \cite{kraskov2004estimating}. It should be noted that KSG is a non-parametric estimation method of the MI and is based on the k-th nearest neighbor. For more details on the KSG the readers is referred to~\cite{kraskov2004estimating}. Results of the privacy-cost trade-off are presented in Fig. \ref{MI}. This confirms that the DD$\mathrm{Q}$L algorithm can reduce the information leakage by incurring a higher energy cost.

\section{Discussion and Concluding Remarks} \label{sec:conclusion}

We examined the privacy-cost trade-off for the SMs system equipped with a rechargeable battery (RB). The optimization problem was formulated as a Markov decisions problem (MDP) where in the reward function the additional cost due to using RB was included while privacy was measured as the average relative distance of the grid load from a constant load. A model-free deep reinforcement learning algorithm, called deep double $\mathrm{Q}$-learning (DD$\mathrm{Q}$L) was used to tackle the problem and was compared with state of the art, classical $\mathrm{Q}$-learning (C$\mathrm{Q}$L). The results using actual SM data showed that the DD$\mathrm{Q}$L algorithm outperforms the C$\mathrm{Q}$L method. In addition, the control of privacy attained by the DD$\mathrm{Q}$L was evaluated in two concrete case studies and in a general information-theoretic sense. Future work will focus on extending our method to accommodate several resources and studying other privacy measures (e.g. MI) to enhance performance.


\section*{Acknowledgment}
This work was supported by Hydro-Quebec, the Natural Sciences and Engineering Research Council of Canada, and McGill University in the framework of the NSERC/Hydro-Quebec Industrial Research Chair in Interactive Information Infrastructure for the Power Grid (IRCPJ406021-14). This project has received funding from the European Union’s Horizon 2020 research and innovation programme under the Marie Skłodowska-Curie grant agreement No 792464.


\bibliographystyle{ieeetr}
\bibliography{HREF}

\begin{thebibliography}{10}

\bibitem{giaconi2018privacy}
G.~Giaconi, D.~Gunduz, and H.~V. Poor, ``Privacy-aware smart metering: Progress
  and challenges,'' {\em IEEE Signal Processing Magazine}, vol.~35, no.~6,
  pp.~59--78, 2018.

\bibitem{sankar2013smart}
L.~{Sankar}, S.~R. {Rajagopalan}, S.~{Mohajer}, and H.~V. {Poor}, ``Smart meter
  privacy: A theoretical framework,'' {\em IEEE Transactions on Smart Grid},
  vol.~4, pp.~837--846, June 2013.

\bibitem{yang2016evaluation}
H.~Yang, L.~Cheng, and M.~C. Chuah, ``Evaluation of utility-privacy trade-offs
  of data manipulation techniques for smart metering,'' in {\em 2016 IEEE
  Conference on Communications and Network Security (CNS)}, pp.~396--400, IEEE,
  2016.

\bibitem{shateri2019deep}
M.~Shateri, F.~Messina, P.~Piantanida, and F.~Labeau, ``Deep directed
  information-based learning for privacy-preserving smart meter data release,''
  in {\em 2019 IEEE SmartGridComm}, pp.~1--7, IEEE, 2019.

\bibitem{kalogridis2010privacy}
G.~Kalogridis, C.~Efthymiou, S.~Z. Denic, T.~A. Lewis, and R.~Cepeda, ``Privacy
  for smart meters: Towards undetectable appliance load signatures,'' in {\em
  2010 First IEEE International Conference on Smart Grid Communications},
  pp.~232--237, IEEE, 2010.

\bibitem{gomez2014smart}
J.~Gomez-Vilardebo and D.~G{\"u}nd{\"u}z, ``Smart meter privacy for multiple
  users in the presence of an alternative energy source,'' {\em IEEE
  Transactions on Information Forensics and Security}, vol.~10, no.~1,
  pp.~132--141, 2014.

\bibitem{giaconi2017smart}
G.~Giaconi, D.~G{\"u}nd{\"u}z, and H.~V. Poor, ``Smart meter privacy with
  renewable energy and an energy storage device,'' {\em IEEE Transactions on
  Information Forensics and Security}, vol.~13, no.~1, pp.~129--142, 2017.

\bibitem{sun2017smart}
Y.~Sun, L.~Lampe, and V.~W. Wong, ``Smart meter privacy: Exploiting the
  potential of household energy storage units,'' {\em IEEE Internet of Things
  Journal}, vol.~5, no.~1, pp.~69--78, 2017.

\bibitem{li2018information}
S.~Li, A.~Khisti, and A.~Mahajan, ``Information-theoretic privacy for smart
  metering systems with a rechargeable battery,'' {\em IEEE Transactions on
  Information Theory}, vol.~64, no.~5, pp.~3679--3695, 2018.

\bibitem{erdemir2019privacy}
E.~Erdemir, P.~L. Dragotti, and D.~G{\"u}nd{\"u}z, ``Privacy-cost trade-off in
  a smart meter system with a renewable energy source and a rechargeable
  battery,'' in {\em ICASSP 2019-2019 IEEE International Conference on
  Acoustics, Speech and Signal Processing (ICASSP)}, IEEE, 2019.

\bibitem{sutton2018reinforcement}
R.~Sutton and A.~Barto, {\em Reinforcement Learning: An Introduction}.
\newblock Adaptive Computation and Machine Learning series, MIT Press, 2018.

\bibitem{szepesvari1998asymptotic}
C.~Szepesv{\'a}ri, ``The asymptotic convergence-rate of q-learning,'' in {\em
  Advances in Neural Information Processing Systems}, 1998.

\bibitem{volodymyr2015human}
M.~Volodymyr, K.~Koray, S.~David, A.~R. Andrei, and V.~Joel, ``Human-level
  control through deep reinforcement learning,'' {\em Nature}, vol.~518,
  no.~7540, pp.~529--533, 2015.

\bibitem{franccois2018introduction}
V.~Fran{\c{c}}ois-Lavet, P.~Henderson, R.~Islam, M.~G. Bellemare, J.~Pineau,
  {\em et~al.}, ``An introduction to deep reinforcement learning,'' {\em
  Foundations and Trends{\textregistered} in Machine Learning}, vol.~11,
  no.~3-4, pp.~219--354, 2018.

\bibitem{mocanu2018line}
E.~Mocanu, D.~C. Mocanu, P.~H. Nguyen, A.~Liotta, M.~E. Webber, M.~Gibescu, and
  J.~G. Slootweg, ``On-line building energy optimization using deep
  reinforcement learning,'' {\em IEEE Tr. on Smart Grid}, 2018.

\bibitem{mnih2015human}
V.~Mnih, K.~Kavukcuoglu, D.~Silver, A.~A. Rusu, J.~Veness, M.~G. Bellemare,
  A.~Graves, M.~Riedmiller, A.~K. Fidjeland, G.~Ostrovski, {\em et~al.},
  ``Human-level control through deep reinforcement learning,'' {\em Nature},
  vol.~518, no.~7540, p.~529, 2015.

\bibitem{hasselt2010double}
H.~V. Hasselt, ``Double q-learning,'' in {\em Advances in Neural Information
  Processing Systems}, pp.~2613--2621, 2010.

\bibitem{van2016deep}
H.~Van~Hasselt, A.~Guez, and D.~Silver, ``Deep reinforcement learning with
  double q-learning,'' in {\em 30th AAAI conference on artificial
  intelligence}, 2016.

\bibitem{beckel2014eco}
C.~Beckel, W.~Kleiminger, R.~Cicchetti, T.~Staake, and S.~Santini, ``The eco
  data set and the performance of non-intrusive load monitoring algorithms,''
  in {\em Proceedings of the 1st ACM Conference on Embedded Systems for
  Energy-Efficient Buildings}, pp.~80--89, ACM, 2014.

\bibitem{kraskov2004estimating}
A.~Kraskov, H.~St{\"o}gbauer, and P.~Grassberger, ``Estimating mutual
  information,'' {\em Physical review E}, vol.~69, no.~6, p.~066138, 2004.

\end{thebibliography}


@article{sun2017smart,
  title={Smart meter privacy: Exploiting the potential of household energy storage units},
  author={Sun, Yanan and Lampe, Lutz and Wong, Vincent WS},
  journal={IEEE Internet of Things Journal},
  volume={5},
  number={1},
  pages={69--78},
  year={2017},
  publisher={IEEE}
}

@article{li2018information,
  title={Information-theoretic privacy for smart metering systems with a rechargeable battery},
  author={Li, Simon and Khisti, Ashish and Mahajan, Aditya},
  journal={IEEE Transactions on Information Theory},
  volume={64},
  number={5},
  pages={3679--3695},
  year={2018},
  publisher={IEEE}
}


@inproceedings{erdemir2019privacy,
  title={Privacy-cost trade-off in a smart meter system with a renewable energy source and a rechargeable battery},
  author={Erdemir, Ecenaz and Dragotti, Pier Luigi and G{\"u}nd{\"u}z, Deniz},
  booktitle={ICASSP 2019-2019 IEEE International Conference on Acoustics, Speech and Signal Processing (ICASSP)},
  year={2019},
  organization={IEEE}
}

@article{giaconi2018privacy,
  title={Privacy-aware smart metering: Progress and challenges},
  author={Giaconi, Giulio and Gunduz, Deniz and Poor, H Vincent},
  journal={IEEE Signal Processing Magazine},
  volume={35},
  number={6},
  pages={59--78},
  year={2018},
  publisher={IEEE}
}

@article{watkins1989learning,
  title={Learning from delayed rewards},
  author={Watkins, CJCH},
  journal={PhD thesis, King's College, University of Cambridge},
  year={1989}
}

@article{sutton2011reinforcement,
  title={Reinforcement learning: An introduction},
  author={Sutton, Richard S and Barto, Andrew G},
  year={2011},
  journal={Cambridge, MA: MIT Press}
}

@book{sutton2018reinforcement,
  title={Reinforcement Learning: An Introduction},
  author={Sutton, R.S. and Barto, A.G.},
  isbn={9780262039246},
  lccn={2018023826},
  series={Adaptive Computation and Machine Learning series},
  url={https://books.google.ca/books?id=6DKPtQEACAAJ},
  year={2018},
  publisher={MIT Press}
}

@inproceedings{szepesvari1998asymptotic,
  title={The asymptotic convergence-rate of Q-learning},
  author={Szepesv{\'a}ri, Csaba},
  booktitle={Advances in Neural Information Processing Systems},
  year={1998}
}

@article{franccois2018introduction,
  title={An introduction to deep reinforcement learning},
  author={Fran{\c{c}}ois-Lavet, Vincent and Henderson, Peter and Islam, Riashat and Bellemare, Marc G and Pineau, Joelle and others},
  journal={Foundations and Trends{\textregistered} in Machine Learning},
  volume={11},
  number={3-4},
  pages={219--354},
  year={2018},
  publisher={Now Publishers, Inc.}
}

@article{mnih2013playing,
  title={Playing atari with deep reinforcement learning},
  author={Mnih, Volodymyr and Kavukcuoglu, Koray and Silver, David and Graves, Alex and Antonoglou, Ioannis and Wierstra, Daan and Riedmiller, Martin},
  journal={NIPS ’13 Workshop on Deep Learning},
  year={2013}
}

@article{volodymyr2015human,
  title={Human-level control through deep reinforcement learning},
  author={Volodymyr, Mnih and Koray, Kavukcuoglu and David, Silver and Andrei, A Rusu and Joel, Veness},
  journal={Nature},
  volume={518},
  number={7540},
  pages={529--533},
  year={2015}
}

@article{mocanu2018line,
  title={On-line building energy optimization using deep reinforcement learning},
  author={Mocanu, Elena and Mocanu, Decebal Constantin and Nguyen, Phuong H and Liotta, Antonio and Webber, Michael E and Gibescu, Madeleine and Slootweg, Johannes G},
  journal={IEEE Transactions on Smart Grid},
  year={2018},
  publisher={IEEE}
}

@article{mnih2015human,
  title={Human-level control through deep reinforcement learning},
  author={Mnih, Volodymyr and Kavukcuoglu, Koray and Silver, David and Rusu, Andrei A and Veness, Joel and Bellemare, Marc G and Graves, Alex and Riedmiller, Martin and Fidjeland, Andreas K and Ostrovski, Georg and others},
  journal={Nature},
  volume={518},
  number={7540},
  pages={529},
  year={2015},
  publisher={Nature Publishing Group}
}


@inproceedings{beckel2014eco,
title={The ECO data set and the performance of non-intrusive load monitoring algorithms},
author={Beckel, Christian and Kleiminger, Wilhelm and Cicchetti, Romano and Staake, Thorsten and Santini, Silvia},
booktitle={Proceedings of the 1st ACM Conference on Embedded Systems for Energy-Efficient Buildings},
pages={80--89},
year={2014},
organization={ACM}
}


@inproceedings{sun2016ev,
  title={EV-assisted battery load hiding: A Markov decision process approach},
  author={Sun, Yanan and Lampe, Lutz and Wong, Vincent WS},
  booktitle={2016 IEEE International Conference on Smart Grid Communications (SmartGridComm)},
  pages={160--166},
  year={2016},
  organization={IEEE}
}

@inproceedings{van2016deep,
  title={Deep reinforcement learning with double q-learning},
  author={Van Hasselt, Hado and Guez, Arthur and Silver, David},
  booktitle={30th AAAI conference on artificial intelligence},
  year={2016}
}

@inproceedings{hasselt2010double,
  title={Double Q-learning},
  author={Hasselt, Hado V},
  booktitle={Advances in Neural Information Processing Systems},
  pages={2613--2621},
  year={2010}
}










@inproceedings{kalogridis2010privacy,
  title={Privacy for smart meters: Towards undetectable appliance load signatures},
  author={Kalogridis, Georgios and Efthymiou, Costas and Denic, Stojan Z and Lewis, Tim A and Cepeda, Rafael},
  booktitle={2010 First IEEE International Conference on Smart Grid Communications},
  pages={232--237},
  year={2010},
  organization={IEEE}
}


@inproceedings{yao2013privacy,
  title={On the privacy-cost tradeoff of an in-home power storage mechanism},
  author={Yao, Jiyun and Venkitasubramaniam, Parv},
  booktitle={2013 51st Annual Allerton Conference on Communication, Control, and Computing (Allerton)},
  pages={115--122},
  year={2013},
  organization={IEEE}
}

@article{gomez2014smart,
  title={Smart meter privacy for multiple users in the presence of an alternative energy source},
  author={Gomez-Vilardebo, Jesus and G{\"u}nd{\"u}z, Deniz},
  journal={IEEE Transactions on Information Forensics and Security},
  volume={10},
  number={1},
  pages={132--141},
  year={2014},
  publisher={IEEE}
}

@article{giaconi2017smart,
  title={Smart meter privacy with renewable energy and an energy storage device},
  author={Giaconi, Giulio and G{\"u}nd{\"u}z, Deniz and Poor, H Vincent},
  journal={IEEE Transactions on Information Forensics and Security},
  volume={13},
  number={1},
  pages={129--142},
  year={2017},
  publisher={IEEE}
}

@ARTICLE{sankar2013smart,
author={L. {Sankar} and S. R. {Rajagopalan} and S. {Mohajer} and H. V. {Poor}},
journal={IEEE Transactions on Smart Grid},
title={Smart Meter Privacy: A Theoretical Framework},
year={2013},
volume={4},
number={2},
pages={837-846},
keywords={distortion;Gaussian processes;hidden Markov models;smart meters;smart meter privacy;theoretical framework;end-user privacy;smart meter measurements;data usage;hidden Markov model;privacy-utility tradeoff problem;stationary Gaussian model;electricity load;mean-square distortion;optimal privacy-preserving solution;implicit distortion noise;frequency components;smart meter privacy;Home appliances;Hidden Markov models;Privacy;Data privacy;Load modeling;Distortion measurement;Batteries;Inference;leakage;privacy;rate-distortion;smart meter;utility},
doi={10.1109/TSG.2012.2211046},
ISSN={1949-3053},
month={June},}

@inproceedings{shateri2019deep,
  title={Deep Directed Information-Based Learning for Privacy-Preserving Smart Meter Data Release},
  author={Shateri, Mohammadhadi and Messina, Francisco and Piantanida, Pablo and Labeau, Fabrice},
  booktitle={2019 IEEE International Conference on Communications, Control, and Computing Technologies for Smart Grids (SmartGridComm)},
  pages={1--7},
  year={2019},
  organization={IEEE}
}

@inproceedings{efthymiou2010smart,
  title={Smart grid privacy via anonymization of smart metering data},
  author={Efthymiou, Costas and Kalogridis, Georgios},
  booktitle={2010 First IEEE International Conference on Smart Grid Communications},
  pages={238--243},
  year={2010},
  organization={IEEE}
}

@inproceedings{yang2016evaluation,
  title={Evaluation of utility-privacy trade-offs of data manipulation techniques for smart metering},
  author={Yang, Huan and Cheng, Liang and Chuah, Mooi Choo},
  booktitle={2016 IEEE Conference on Communications and Network Security (CNS)},
  pages={396--400},
  year={2016},
  organization={IEEE}
}

@article{kraskov2004estimating,
  title={Estimating mutual information},
  author={Kraskov, Alexander and St{\"o}gbauer, Harald and Grassberger, Peter},
  journal={Physical review E},
  volume={69},
  number={6},
  pages={066138},
  year={2004},
  publisher={APS}
}

\end{document}